\documentstyle[preprint,aps,epsf]{revtex}

\begin{document}

\draft

\title{Quantum Critical Dynamics of the
Random Transverse Field Ising Spin Chain}

\author{Heiko Rieger}
\address{HLRZ, Forschungszentrum J\"ulich, 52425 J\"ulich, Germany}

\author{Ferenc Igl\'oi}
\address{Research Institute for Solid State Physics, 
H-1525 Budapest, P.O.Box 49, Hungary$^*$\\
Institut f\"ur Theoretische Physik, Universit\"at zu
K\"oln, 50937 K\"oln, Germany}

\date{April 16, 1997}

\maketitle

\begin{abstract}
Dynamical correlations of the spin and the energy density are
investigated in the critical region of the random transverse-field
Ising chain by numerically exact calculations in large finite systems
$(L \le 128)$. The spin-spin autocorrelation function is found to
decay proportional to $(\log\,t)^{-2x_m}$ and $(\log\,t)^{-2x_m^s}$ in
the bulk and on the surface, respectively, with $x_m$ and $x_m^s$ the
bulk and surface magnetization exponents, respectively.  On the
other hand the critical energy-energy autocorrelation functions have a
power law decay, which are characterized by novel critical exponents
$\eta_e \approx 2.2$ in the bulk and $\eta_e^s \approx 2.5$ at the
surface, respectively. The numerical results are compared with the
predictions of a scaling theory.
\end{abstract}

\pacs{05.50.+q, 64.60.Ak, 68.35.Rh}

\newcommand{\bc}{\begin{center}}
\newcommand{\ec}{\end{center}}
\newcommand{\be}{\begin{equation}}
\newcommand{\ee}{\end{equation}}
\newcommand{\beqn}{\begin{eqnarray}}
\newcommand{\eeqn}{\end{eqnarray}}

The asymptotic behavior of the time-dependent correlation functions
for interacting many-body systems turned out a very difficult subject
of theoretical research. Exact results in this field are scarce, one
can mention the one-dimensional spin $1/2$ XY-model \cite{McCBA} and
the Ising chain in a transverse-field \cite{perk}.  Both models can be
mapped onto a system of non-interacting fermions, where the
equal-position correlation functions are calculated by the Pfaffian
method utilizing the theory of T\"oplitz determinants.

In this Letter we consider - at the first time - the critical
dynamical correlations of an interacting quantum system in the
presence of quenched (i.e. time-dependent) disorder. It has recently
become clear that quenched disorder has rather different effects on
phase transitions in quantum systems \cite{qpt} than on those
thermally driven phase transitions. For example, in the Griffiths
phase, which is situated at the disordered side of the critical point,
the susceptibility has an essential singularity in classical systems,
whereas in a quantum system the corresponding singularity is stronger,
it is in a power law form.

Here we consider the prototype of random quantum
systems the one-dimensional random transverse-field Ising model
defined by the Hamiltonian:
\be
H=-\sum_l J_l \sigma_l^x \sigma_{l+1}^x-\sum_l h_l \sigma_l^z\;,
\label{hamilton}
\ee
where the $\sigma_l^x,~\sigma_l^z$ are Pauli matrices at site $l$
and the $J_l$ exchange couplings and the $h_l$ transverse-fields
are random variables with distributions $\pi(J)$ and $\rho(h)$,
respectively. The Hamiltonian in (\ref{hamilton}) is closely related to the
transfer matrix of a classical two-dimensional layered Ising model,
which was first introduced and studied by McCoy and Wu \cite{mccoy}.

The static critical behavior of the random transverse-field Ising
model in (\ref{hamilton}) has been studied analytically and
numerically by several authors
\cite{fisher,young,mckenzie,profiles}. The system possesses a critical
point at $\delta=[\ln J]_{\rm av}-[\ln h]_{\rm av}=0$, and has a
spontaneous ferromagnetic order if the average couplings are stronger
than the average fields.  (We use the bracket $[\dots]_{\rm av}$ to
denote disorder averages.) The critical properties of the model, which
are known through exact and conjectured results to a large extent, are
in many respects different from that of pure systems. One important
difference, that in the random system - due to a broad distribution of
various physical quantities - the typical and average quantities are
usually different and the rare events dominate the critical
properties. For instance the static average spin-spin correlation
function is expected to behave as
\be
G^m_l(r)=\left[\langle\sigma_l^x\sigma_{l+r}^x\rangle\right]_{\rm av}
={1 \over r^{2 x_m}} \exp(-r/\xi)\;,
\label{static}
\ee
where $\langle\dots\rangle$ means the (zero-temperature) expectation
value.  For the random transverse-field Ising model the average
correlation length $\xi \sim \delta^{-\nu}$ diverges with the true
exponent $\nu=2$ and the decay exponent $x_m=1-\omega/2\approx.191$ is
expressed in terms of the golden mean $\omega=(1+\sqrt{5})/2$. The
decay of the average end-to-end distance critical correlations
involves the surface magnetization exponent $x_m^s=1/2$. On the other
hand the {\it typical} correlation length diverges with $\nu_{\rm
typ}=1$ and the {\it typical} critical correlations are of a stretched
exponential form: $-\log G^m_{\rm typ}(r) \sim r^{1/2}$.  In
contrast the critical energy-density correlation function
$G^e_l(r)=\left[\langle\sigma_l^z\sigma_{l+r}^z\rangle\right]_{\rm
av}$ is a self-averaging quantity and at the critical point it behaves
as $-\log G^e(r) \sim r^{1/2}$, like its typical value.

In this Letter we consider the time-dependent correlation functions
\be
G^m_l(r,t)=[\langle\sigma_l^x(t)\sigma_{l+r}^x\rangle]_{\rm av}
\qquad{\rm and}\qquad
G^e_l(r,t)=[\langle\sigma_l^z(t)\sigma_{l+r}^z\rangle]_{\rm av}
\label{energycorr}
\ee
at the critical point, both in the bulk and at the surface of the
system. In a quantum system statics and dynamics are inherently
related and the time evaluation is given via the Heisenberg picture by
$\sigma_l^x(t)=\exp(tH) \sigma_l^x \exp(-tH)$.  For simplicity here we
confine ourselves to the {\it auto}correlations, i.e.\ $r=0$,
dynamical two-site correlations will be discussed elsewhere
\cite{bigpaper}.

To start our study we present a scaling framework for the quantum
critical dynamics of the model (\ref{hamilton}). Consider the general
time and position dependent correlation function
$\langle\sigma_l^x(t)\sigma_{l+r}^x\rangle$, which can be written as
\be
\langle\sigma_l^x(t)\sigma_{l+r}^x\rangle=\sum_n
\langle 0|\sigma_l^x|n\rangle\langle n|\sigma_{l+r}^x |0\rangle \exp[-t
(E_n-E_0)]\;.
\label{general}
\ee
Here $|n\rangle$ denotes the $n$-th excited state of $H$ in
eq.\ (\ref{hamilton}) with energy $E_n$.  Before performing the disorder
average we note that this correlation function is not self averaging
at the critical point. To see its scaling behavior at the critical
point we present the following simple argument. The random samples can
be divided into two groups. In the {\it typical} samples (i.e. which
appear with probability one) the critical correlations decay faster
than any power law. On the other hand a vanishing fraction of the
samples (the so called {\it rare events}) is ordered at the critical
point and the correlation function measured on these samples is of
order $O(1)$. The disorder average of the correlation function is then
determined by the rare events and the corresponding scaling behavior
is governed by the scaling properties of the probability distribution
of these rare realizations.

For example the probability $P(l)$, which measures the occurrence of
samples with a finite local magnetization $m(l)=O(1)$ at site $l$
(take for instance fixed boundary conditions, or consider an
off-diagonal matrix element in the case of free b.c.\ , see
\cite{profiles}), scales as the average critical magnetization
$P(l/b)=b^{-x_m} P(l)$, when lengths are rescaled by a factor
$b>1$. For equal time correlations in the rare realizations the local
magnetization is of order $O(1)$ at both spatial coordinates. The
corresponding joint probability distribution $P_2(l,l+r)$ factorizes
for large spatial separations $\lim_{r \to \infty} P_2(l,l+r)=P(l)
P(l+r)$, since the disorder is uncorrelated. Consequently the spatial
correlations follow the scaling rule:
\be
G^m(r,t=0)=b^{-2x_m} G^m(r/b,t=0)\;,
\label{static2}
\ee
whereas for end-to-end distance correlations we have the surface
magnetization scaling dimension $x_m^s$.  Now taking $r=b$ we recover
the known critical decay as given in eq.\ ({\ref{static}).

For critical time-dependent spin-spin autocorrelations, however, the
scaling behavior is different from that in eq.\ (\ref{static2}). This
is due to the fact that the disorder is strictly correlated along the
time axis and the probability for the occurence of a rare sample with
$m(l)=O(1)$ at different times is simply $P_2((l,t),(l,0)) \sim
P(l)$. Thus the scaling behavior of the critical magnetization
autocorrelation function satisfies the scaling rule:
\be
G^m(r=0,\ln t)=b^{-x_m} G^m(r=0,\ln t/b^{1/2})\;,
\label{spindyn1}
\ee
where we have made use of the relation between the relevant time $t_r$
and length $\xi$ scales $\sqrt{\xi}\sim\ln\,t_r$ \cite{fisher,young}.
Note that the usual scaling combination is $t/b^z$, however, the
critical dynamical exponent $z$ is $\infty$ here.  Taking now the
length scale as $b=(\ln t)^2$, we obtain
\be
G_m(r=0,t) \sim (\ln t)^{-2x_m}
\label{spindyn2}
\ee
For the surface autocorrelation function the scaling relation in
eq.\ (\ref{spindyn1},\ref{spindyn2}) and consequently the decay exponent
involves the surface magnetization exponent $x_m^s$.

For energy density autocorrelations the typical realizations govern
the scaling properties at the critical point. The relevant quantity is
now the matrix-element \hfill $[|\langle
0|\sigma_l^z|n\rangle|^2]_{\rm av}$ on the r.h.s. of eq.\
(\ref{general}), which scales in an exponential form: $\log [|\langle
0|\sigma_l^z|n\rangle|^2]_{\rm av}=b^{-1/2} \log [|\langle
0|\sigma_{l/b}^z|n\rangle|^2]_{\rm av}$ \cite{profiles}.  Consequently
the critical energy density autocorrelations satisfy the scaling
relation:
\be
\log G^e(r=0,\ln t)=b^{-1/2} \log G^e(r=0,\ln t/b^{1/2})\;,
\label{energydyn}
\ee
and with $b=(\ln t)^2$ one obtains a power law dependence of $G^e(r=0,t)$
with novel, non-trivial exponents:
\be
G^e(r=0,t)\sim t^{-\eta_e}\;.
\label{energydyn1}
\ee

In the actual calculations we transformed the model in eq.\ 
(\ref{hamilton}) into a free fermion model \cite{fermion}, where the
correlation functions are expressed by averages of fermion operators,
which are then calculated by Wick's theorem and by the Pfaffian method
\cite{pfaffian}. We use free boundary conditions, in which case the
most convenient representation is given in \cite{igloiturban}, which
necessitates only the diagonalization of an $2L \times 2L$ matrix.
From the corresponding eigenvalues and eigenvectors one obtains the
elements of the Pfaffian, which is then evaluated by calculating the
determinant of the corresponding antisymmetric matrix. Details of the
calculations will be presented elsewhere \cite{bigpaper}.

The critical properties of the random quantum spin chains are expected
to be independent of the details of the distributions of the couplings
and the fields. In this Letter we consider the binary distribution
$\pi(J)={1 \over 2} \delta(J-\lambda) + {1 \over 2} \delta(J-\lambda^{-1})$
and $h=h_0$, and the uniform distribution $
\pi(J)=\Theta (1-J) \Theta (J)$ and 
$\rho(h)=h_0^{-1} \Theta (h_0-h)\Theta (h)$.
%
%
%
%
In both cases the critical point is at $h_0=1$.  All numerical data
which we present below are averaged over $50000$ samples.

First we study the critical spin-spin autocorrelation function for
imaginary times $t=-i \tau$ in the bulk (i.e.\ at the site $i=L/2$)
and at the surface (i.e.\ at site $i=1$). As shown in Fig.\ 1a the
finite lattice results fall to the same curve for $\log \tau \le
\sqrt{L}$ and the critical temporal decay takes place on a logarithmic
scale $G^m_{L/2}(\tau) \sim ( \log \tau )^{-2 x_m}$ in agreement with
the scaling prediction (\ref{spindyn2}). For surface correlations the
numerical calculation is less demanding and one can go up to finite
systems of size $L=128$. As can be seen in Fig.\ 1b in this case the
logarithmic decay depends on the surface magnetization exponent:
$G_1^m(\tau) \sim ( \log \tau )^{-2 x_m^s}$.

The autocorrelation functions in real time generally have an oscillatory
character. In the random system the average over different oscillating
functions results in a complicated looking behaviour, as we demonstrate it
for the surface autocorrelation function in Fig.\ 2a. Its Fourier transform,
however, has a nice scaling character. We actually consider
\be
\chi_1^m(\omega)
=\frac{1}{2\pi}\int_{-\infty}^{\infty} dt\,e^{i\omega t}\,
\int_{-\infty}^{\infty} d\tau\,G_1^m(t+i\tau)
=\frac{2}{\omega}\,|\langle\omega|\sigma_1^x|0\rangle|^2\;,
\label{fourier}
\ee
where $\langle\omega|$ is a state with an excitation energy 
$E_{\rm exc.}-E_0=\omega$. For small frequencies $\omega$ we expect
the finite size scaling form of $\chi_1^m(\omega)$ to be given by
\be
\chi_1^m(\omega,L)\sim
\omega^{-1}\,L^{-1}\,\tilde{\chi}( \log(\omega)/L^{1/2} )
\ee
with the scaling combination $\log(\omega)/L^{1/2}$ replacing
$\log(t)/L^{1/2}$ from (\ref{spindyn1}).  In Fig.\ 2b we show a
corresponding scaling plot that yields a good data collapse.

Next we turn to analyze the energy density autocorrelation function at
the critical point. As seen on Fig.\ 3a the energy density
autocorrelation function is described by a power law dependence in
imaginary time $\tau$ as $G_{L/2}^e(\tau) \sim \tau^{-\eta_e}$ in
agreement with the scaling prediction (\ref{energydyn}) and
(\ref{energydyn1}).  The decay exponent $\eta_e \simeq 2.2$ is
universal, i.e. it does not depend on the type of the
randomness. A similar power law decay is found for the surface energy
autocorrelations in Fig.\ 3b, with a surface critical exponent $\eta_e^s
\simeq 2.5$. These novel critical exponents complete our knowledge
about the critical behavior of the random transverse-field Ising spin
chain.

To summarize we have studied dynamical correlations at the critical point
of the random transverse-field Ising spin chain. We showed that the
magnetization autocorrelation function has anomalous logarithmic decay,
whereas the energy-density autocorrelations decay as a power law with
novel critical exponents. There are still many interesting aspects of
the dynamical behavior of random quantum systems. Here we mention the
dynamical properties in the Griffiths phase, the temperature dependent
autocorrelations and the dynamical two-site correlations. The study of
these and other related problems are in progress \cite{bigpaper}.

Acknowledgment: This study was partially performed during F.\ I.'s
visit in K\"oln. This work has been supported by the Hungarian
National Research Fund under grants No OTKA TO12830 and OTKA TO23642,
and the Sonderforschungsbereich (SFB) 341 (K\"oln--Aachen--J\"ulich).
H.\ R.'s work was supported by the Deutsche Forschungsgemeinschaft
(DFG).

\begin{figure}
\epsfbox{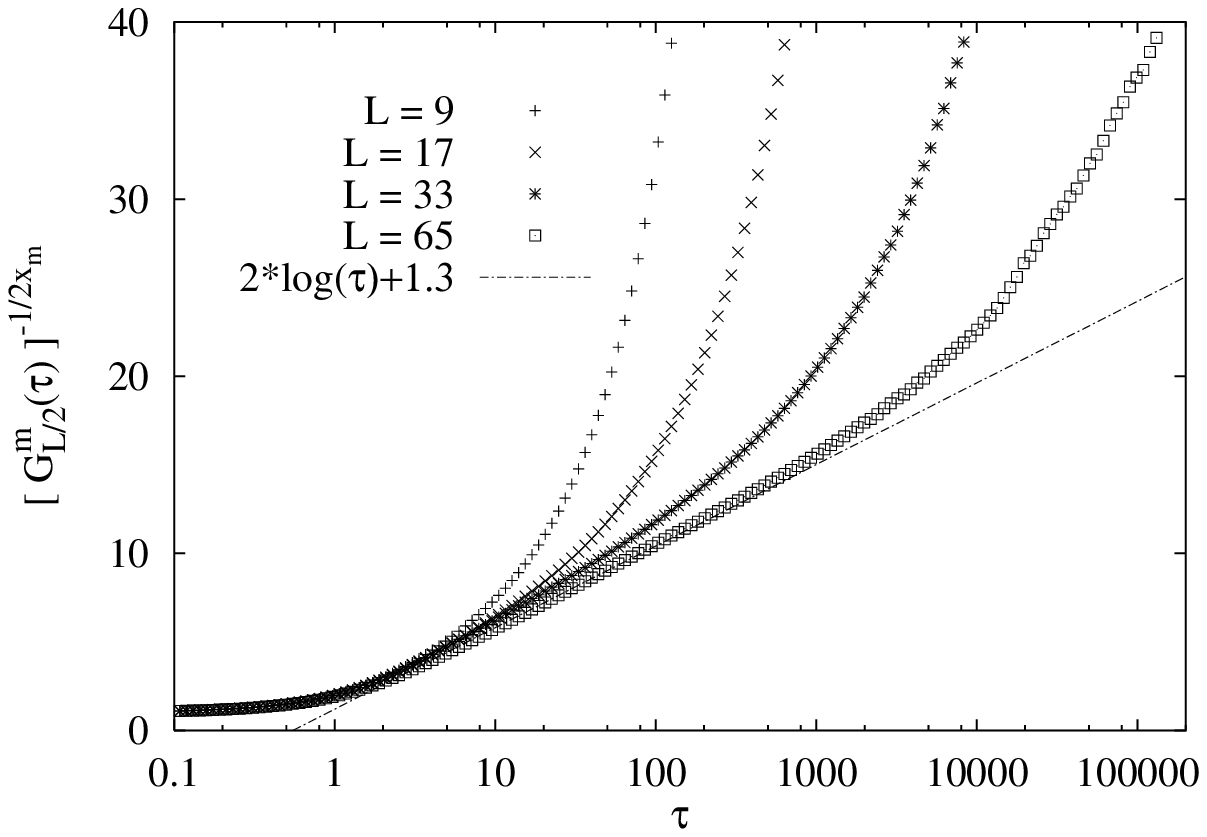}
  {\bf FIG.\ 1a:}
  Bulk spin-spin autocorrelation function 
  $G_{L/2}^m(\tau)=[\langle\sigma_{L/2}^x(t)\sigma_{L/2}^x\rangle]_{\rm av}$ 
  in imaginary time for various system sizes (and the uniform
  distribution). Note that we have chosen $L$ to be odd, so that
  $L/2$ denotes the central spin. In this plot with 
  $[ G_{L/2}^m(\tau) ]^{-1/2x_m}$ on
  linear scale versus $\tau$ on a logarithmic scale the infinite
  system size limit is expected to lay on a straight line as
  indicated.
\end{figure}

\begin{figure}
\epsfbox{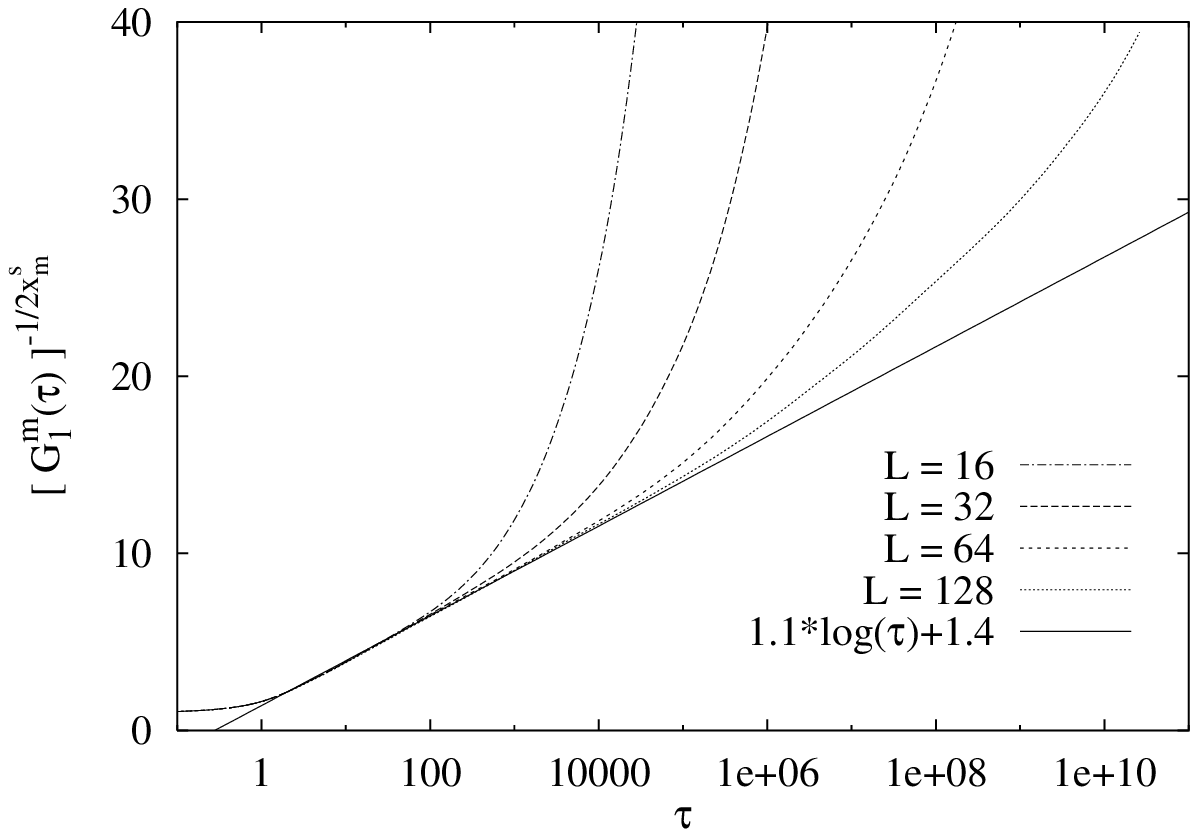}
  {\bf FIG.\ 1b:}
  Same as a) for the surface spin-spin autocorrelation function 
  $G_1^m(\tau)=[\langle\sigma_1^x(\tau)\sigma_1^x\rangle]_{\rm av}$ 
  in imaginary time.
\end{figure}
\vfill
\eject

\begin{figure}
\epsfbox{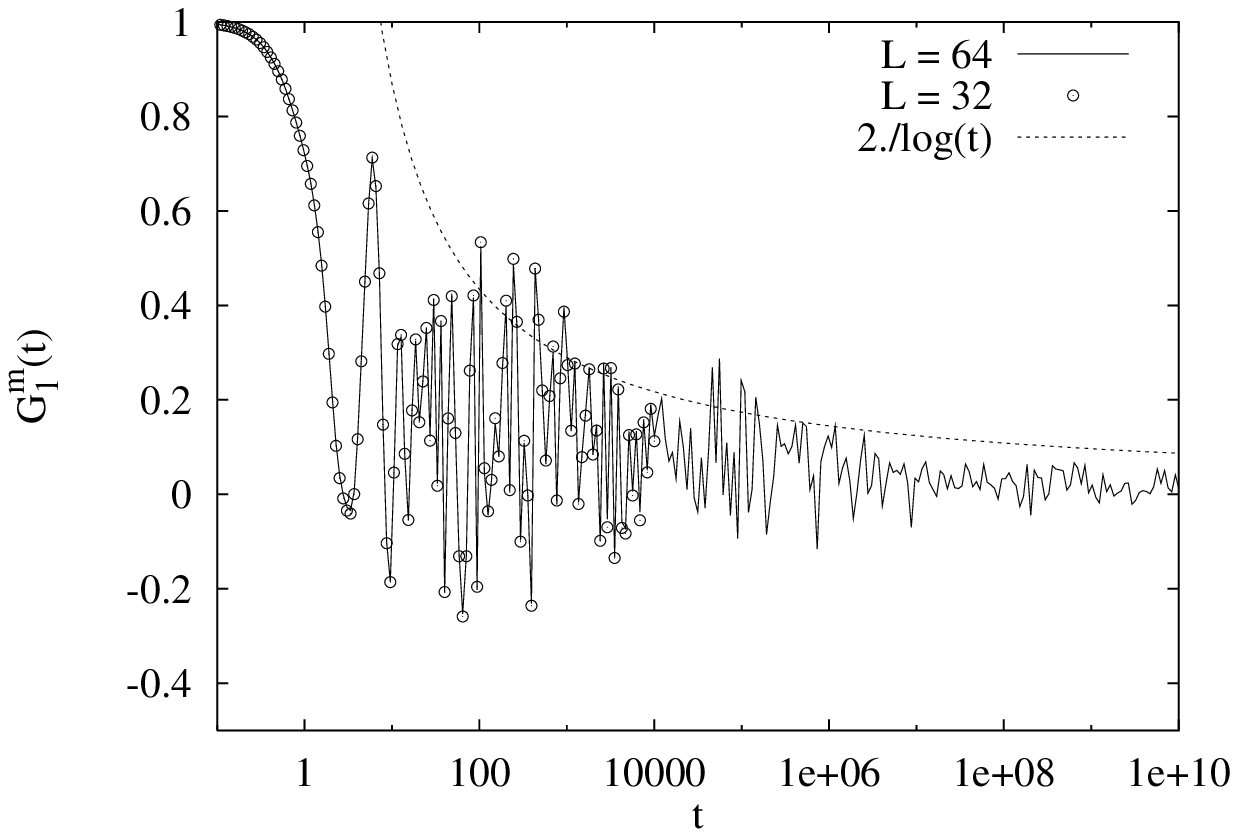}
\epsfbox{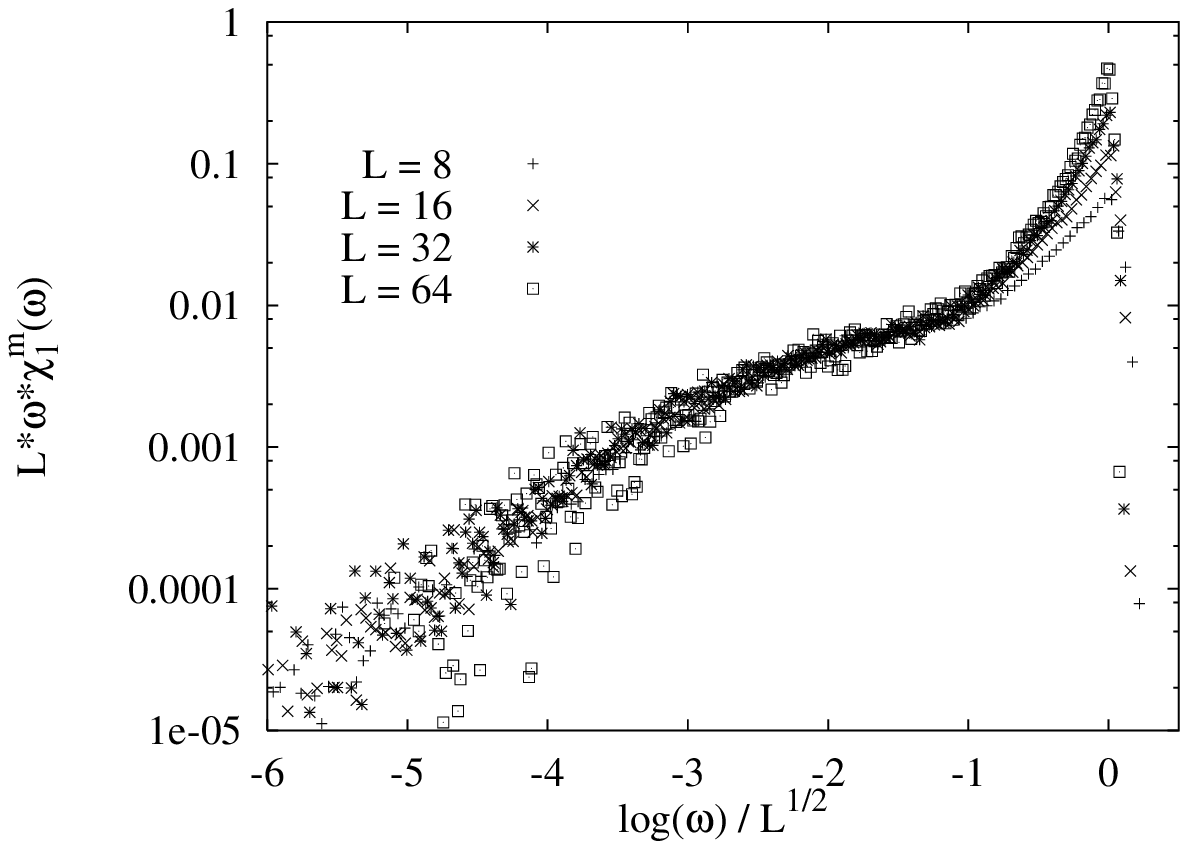}
  \
  {\bf FIG.\ 2a:} (Top) Surface spin-spin autocorrelation function
  $G_1^m(t)$ in real time for the binary distribution with
  $\lambda=4$. The data for L=64 and those shown for L=32 are {\it
  exactly} identical, although both data sets have different disorder
  realization. The expected $1/\log(t)$ behavior for the envelope
  indicated by the broken line is only a guide to the eye.
  \protect{\hfill\newline}  
  {\bf FIG.\ 2b:} (Bottom) Scaling plot of the Fourier transformed 
  surface spin-spin autocorrelation function $\chi_1^m(\omega)$
  (\protect{\ref{fourier}}) for the binary distribution and
  $\lambda=4$.
\end{figure}
\vfill
\eject

\begin{figure}
\epsfbox{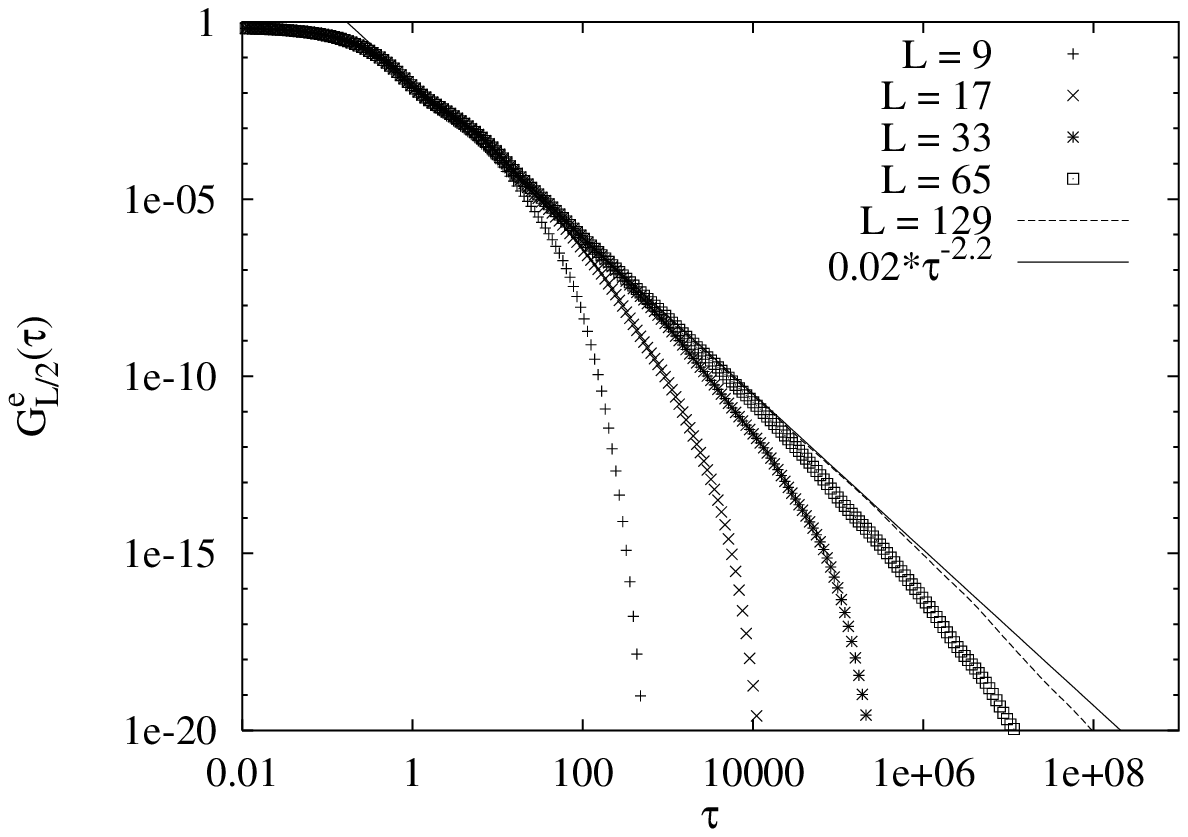}
\epsfbox{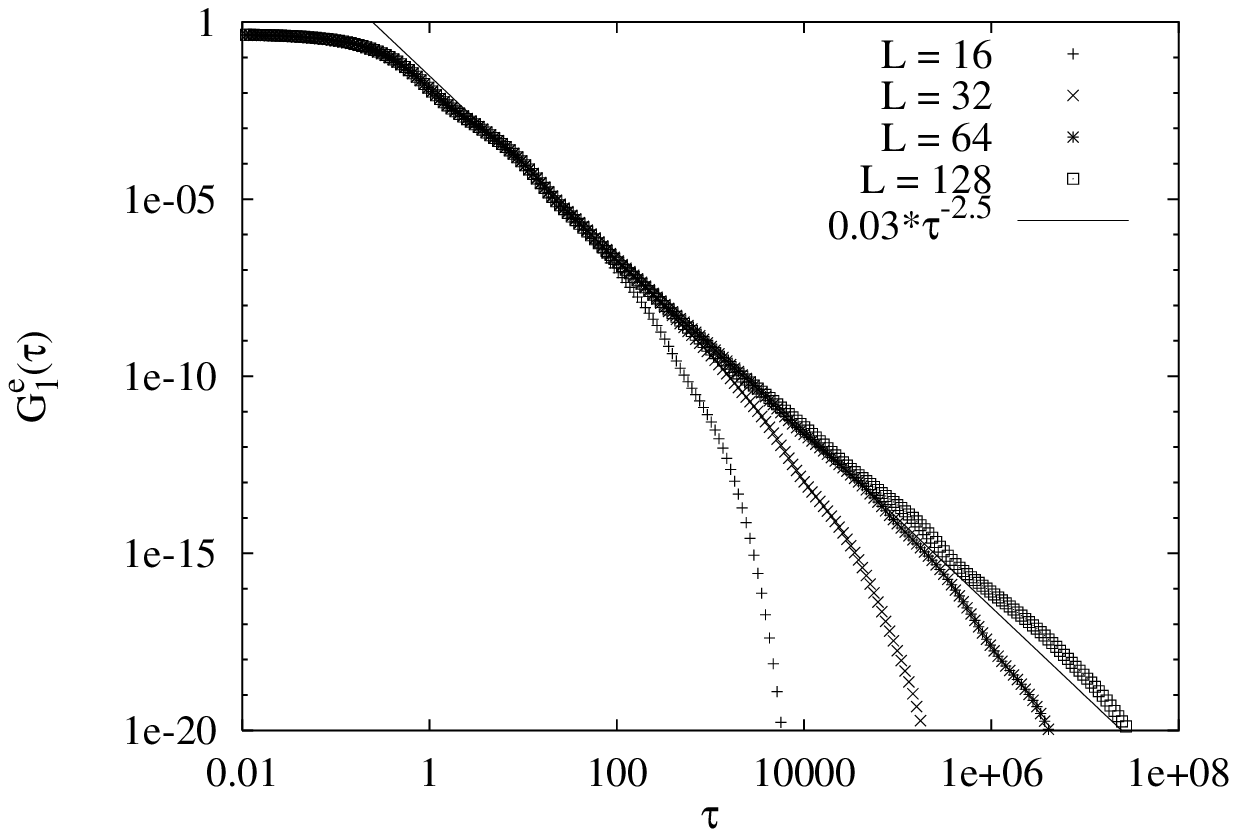}
  {\bf FIG.\ 3a:} (Top):
  Bulk energy-energy autocorrelation function 
  $G_{L/2}^e(\tau)=[\langle\sigma_{L/2}^z(\tau)\sigma_{L/2}^z\rangle]_{\rm av}$ 
  in imaginary time for various system sizes (and the binary
  distribution, $\lambda=4$) in a log-log plot. The straight line has
  slope $-2.2$, which yields our estimate for the exponent $\eta_e$.
  {\bf FIG.\ 3b:} (Bottom)
  Same as a) for the surface energy-energy autocorrelation function 
  $G_1^e(\tau)=[\langle\sigma_1^z(\tau)\sigma_1^z\rangle]_{\rm av}$ 
  in imaginary time. The straight line has
  slope $-2.5$, which yields our estimate for the exponent $\eta_e^s$.
\end{figure}

\end{document}